\documentclass[12pt]{article}
\usepackage{amsmath,amsfonts,amssymb,a4,epsfig,color,graphics}

\newcommand{\R}{{\mathbb{R}}}
\newcommand{\C}{{\mathbb{C}}}
\newcommand{\Z}{{\mathbb{Z}}}
\newcommand{\N}{{\mathbb{N}}}

\newcommand{\1}{{\bf 1}}

\def\ha{\frac{1}{2}}
 
\def\pa{\partial}
\def\ra{\rightarrow}
\def\preuve{\begin{proof}} 
\def\ga{\alpha}
\def\gb{\beta}

\def\gd{\delta}
\def\ge{\varepsilon}

\def\gl{\lambda}

\def\gr{\rho}

\def\san{San V{\~u} Ng{\d o}c}

\newtheorem{defi}{Definition}

\newtheorem{lemm}{Lemma}

\newtheorem{rem}{Remark}
\newtheorem{coro}{Corollary}
\newtheorem{theo}{Theorem}

\newenvironment{demo}{\noindent {\it Proof.--}
      \begin{quotation}\noindent}{\end{quotation}\hfill$\square $}

\begin{document}

\title{On the remainder in the Weyl formula for
the Euclidean disk}
\author{Yves Colin de Verdi\`ere \footnote{Institut Fourier,
 Unit{\'e} mixte
 de recherche CNRS-UJF 5582,
 BP 74, 38402-Saint Martin d'H\`eres Cedex (France);
yves.colin-de-verdiere@ujf-grenoble.fr}
}


\maketitle

\begin{abstract} We prove  a 2-terms Weyl formula for the
  counting function $N(\mu)$ 
 of the spectrum of the Laplace operator in the
  Euclidean disk with a sharp remainder estimate $O\left(\mu^{2/3}\right)$.

\end{abstract}

\section*{Introduction}
 Let us denote by   $0<\gl_1<\gl_2 \leq \cdots $
 the  eigenvalues for the Dirichlet Laplacian of
some bounded connected smooth domain $X$  in the Euclidean plane.
It has been shown by Ivrii \cite{Iv} that the following 2-terms Weyl formula
holds under some genericity assumption on the periodic orbits
of the associated billiard ball problem:
 if $N_X(\mu )=\# \{ j~|~\gl_j \leq \mu^2 \}$,
\[ N_X(\mu)=\frac{|X|}{4\pi}\mu^2-\frac{|\pa X|}{4\pi}\mu +R(\mu)  \]
with $R(\mu)=o(\mu)$.
Moreover, this result is quite optimal: Lazutkin and Terman  
 \cite{L-T} showed  that there is no 
$\gd >0$ so that  an estimate 
$R(\mu)=O\left(\mu ^{1-\gd }\right)$   holds for
all smooth convex domains.

Our goal is to get an upper bound for  $R(\mu)$
 in the case of the Euclidean disk.
Our main result\footnote{After having completed this work,
we  learned from I. Polterovich that the same
  result
has been announced  in 1964
  by N. V. Kuznecov and  B. V.  Fedosov in \cite{K-F}.
The method is similar to ours. We give here an independent  complete
derivation of the needed Van der Corput result} is:
\begin{theo}\label{theo:main}
\[  N_{\rm disk}(\mu )=\mu ^2/4 -\mu /2 + O\left(\mu^{2/3}\right)~.\]
\end{theo}

 The proof  is based on the
explicit
expression  of the eigenvalues as the squares of the zeros  of the Bessel
functions
$J_n$ as well as on some precise asymptotics of these zeros which goes back
to Olver (see \cite{Ol,CGJ}). This way, we have  to study
 a lattice point problem
in some domain with cusps.
A rather general  lattice point problem  was  studied
by Van der Corput \cite{VdC}, see also 
\cite{Hl,He,Ra,CdV1}. In  \cite{CdV2}, a similar method was used in order to
get a good remainder estimate for some surfaces of revolution. 
Let us note also that the same  remainder estimate holds for
the integrable polygonal billiards like the rectangles
or the equilateral triangles: this is a direct consequence
of the explicit formula for  the eigenvalues which reduces the question
directly to a lattice point problem for which the Van der Corput's
result
applies.

\section{The spectrum of the unit disk}\label{sec:unit}

We consider  the spectrum of the Euclidean Laplacian
$\Delta _{\rm disk}=-\pa _x^2 -\pa_y^2$  in the 
unit disk in $\R^2_{x,y}$ with Dirichlet boundary conditions.
As it is well known and can be checked by separation
of variables, the eigenvalues of $\Delta _{\rm disk}$
are the squares of the zeros of the Bessel functions 
$J_n,~n\in \Z$.
Let us recall that 
 \begin{equation} \label{equ:bessel}
 J_n(x)=\frac{1}{2\pi}\int _{-\pi}^\pi e^{i(x \sin t -nt )} dt
~,\end{equation}
and that  we have the following identities
$J_n(-x)=(-1)^n J_n(x),~J_{-n}(-x)=J_n(x)$.
Let us denote by
$|n| < x_1(n) < x_2(n) < \cdots < x_k(n) < \cdots $
the positive zeros of $J_n$.
Then the spectrum of  $\Delta _{\rm disk}$, with multiplicity, is
given by
\[ \sigma =\{ x_k(n)^2~|~n\in \Z,~ k=1,\cdots,\}  ~.\]
\begin{figure}[hbtp]
  \begin{center}
    \leavevmode
\input{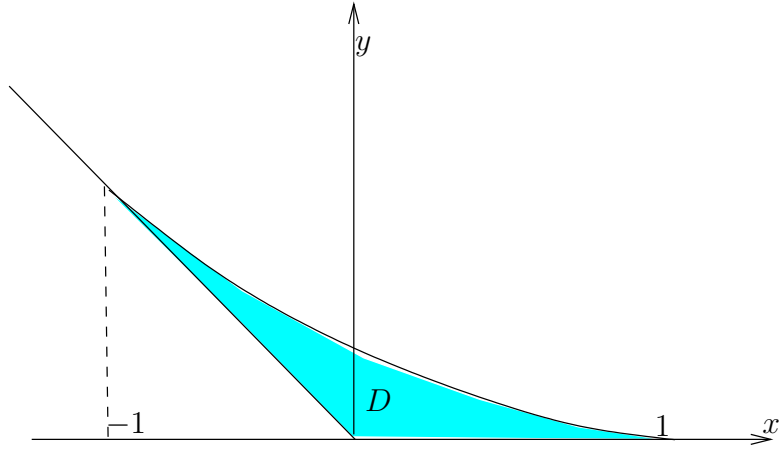}
       \caption{the domain $D$}
    \label{fig:D}
  \end{center}
\end{figure}

In order to describe the asymptotics of the zeros of Bessel
 functions, we introduce the domain
$D$ in $\R^2$  defined by
\[ D= \{ (x,y)~|~-1\leq x \leq 1,~ y\leq g(x),~y \geq \max (0,-x) ~\} \]
with 
\[ g(x)=\frac{1}{\pi} \left( \sqrt{1-x^2 }-x \arccos x
\right) ~.\]

Let us define  ${\cal R} =\{ (n,k-1/4)~|~(n,k)\in \Z^2 \}$ 
and  $S=\{ (x,y)~|~y\geq \max (0,-x)\}$. 
Let  $F:S \ra \R $ be  the function homogeneous of degree $1$
 which satisfies $F\equiv 1$ on the graph of $g$.
The spectrum of the disk is approximately given by
$\{ \gl_j~|~j\in \N  \}\sim \{ F(m)~|~m\in {\cal R}\cap S \}$.
More precisely, 
for $n\geq 0$, $x_k(n)\sim F(n,k-1/4)$, while for $n<0$,
$x_k(n)\sim F(n,k+|n|-1/4)$.
This will reduce our problem to a lattice point problem:
\[ N_{\rm disk}(\mu )\sim N_{\rm D}(\mu)= \# \{ m \in {\cal R} \cap
\mu D \} 
~.\] 
\begin{rem}
Let us note  that $D$ and ${\cal R}$ are invariant by the linear
involution
$J(x,y)=(-x, y +x)  $.
This corresponds to the fact that $J_n$ and $J_{-n} $ have
the same zeros.
\end{rem}

In order to complete the argument, we will have to study the 
lattice point problem (Section \ref{sec:lpp}) and 
  to show how close 
 $ N_{\rm disk}(\mu )$ and $N_{\rm D}(\mu)$ are (Section
 \ref{sec:lpp/disk}).
The
first  part uses the method of Hlawka, Herz and Randol \cite{Hl,He,Ra}
 for studying smooth lattice point
problems and the
second  part is done using Olver's asymptotics 
for the zeros of Bessel functions: we  re-derived it  in \cite{CGJ}
using the integral representation of the Bessel functions
and the general theory of oscillatory integrals associated 
to versal unfoldings of singularities as explained in \cite{G-S}.

\section{A lattice point problem with a cusp}\label{sec:lpp}

Let us denote by ${\cal R} $ the lattice
${\cal R}:=\{ \left(n,k-\gb\right)~|~\left(n,k\right)\in \Z^2 \} $
 with $0<\beta <1$. 
Let us consider a domain $G\subset \R^2$ with a cusp:
 $G=\{\left(x,y\right)~|~0\leq x \leq 1,
~0\leq y \leq g\left(x\right) \} $ with
$g\left(x\right)\sim a \left(1-x\right)^{3/2}$ with $a>0$ near $x=1$.
We consider the weighted  lattice point problem defined by the
counting
function
\[ N_{G,\beta,\chi}\left(\mu\right)=\sum _{m=(m_1,m_2)\in \mu G\cap  {\cal R}}
\chi \left({m_2}/{m_1} \right)  ~,\]
with $\chi \in C_o^\infty (\R)$ with $\chi \equiv 1 $ near $0$ and 
the support of $\chi $ small enough.

We have the following 2-term Weyl estimate:
\begin{theo} \label{theo:lpp} Under the previous assumptions on 
$G$, $\beta$  and $\chi$, we have 
\[ N_{G,\beta,\chi}\left(\mu \right)
=\left(\int _G \chi\left(y/x\right)dx dy \right)\mu^2
+ \left(\gb-\ha   \right)\mu + 0\left(\mu ^{2/3}\right)~.\]
\end{theo}
\begin{coro}\label{cor:lpp} If $D$ is the domain defined in Section
\ref{sec:unit} and  $\beta=1/4$, we have
\[ N_D(\mu )={\rm Area }(D)\mu^2 -\frac{\mu}{2}+ 0\left(\mu ^{2/3}\right)~.\]
\end{coro}
{\it Proof of Corollary \ref{cor:lpp}:}
we decompose $N_D$ into 3 terms: one for each cusp and one inner
term using an homogeneous partition of unity.
The corollary follows from the previous Theorem for the parts near the cusps
and from  the  classical estimates going back at least to
Van der Corput \cite{VdC} (see also \cite{Ra, CdV1}) for the inner
part.
In order to use Van der Corput estimates $O\left(\mu^{2/3}\right)$,
we need to check the strict convexity, in fact  the non vanishing 
of the curvature
of the graph of $g$: this comes from the fact that
$g''(x)=\left(1-x^2\right)^{-\ha}>0 $.\hfill$\square $

{\it Proof of Theorem \ref{theo:lpp}:}
let us denote by $B(m,r)$ the Euclidean ball of center $m$ and
radius $r$. 
 We  can first replace $\chi (y/x)$ by 
the smooth function 
$\chi_0(x,y)=\chi (y/x)(1-\phi (x,y))$ with $\phi \in C_o^\infty $
with support in the ball $B(0,\min(\gb, 1-\gb))$
 and $\equiv 1$ near $0$
because there is no element of ${\cal R}$ in the support
of $\phi $. The smooth function $\chi _0$ is a classical symbol of
degree 
$0$: $\pa ^j _x \pa _y^k \chi _0(x,y) =0( (1+|x|+|y|)^{-(j+k)})$.

Let us give a positive function $\rho \in C_o^\infty
 \left(\R^2\right)$
with ${\rm Support }(\rho)\subset \{ x^2 +y^2 <1 \}$
and $\int _{\R^2}\rho (x,y) dxdy =1$,  define 
$\rho_\ge =\rho \left(./\ge\right)/\ge^2$ with $\ge =\mu^{-1/3}$,
and consider 
\begin{equation} \label{equ:neps}
 N_\ge^\pm  \left(\mu \right) = \sum_{m\in{\cal R}} \left(\chi_0
\1_{G^\pm_{\mu,\ge}
}\star \rho_\ge \right) \left(m\right) \end{equation}
where
 \[   G^+_{\mu,\ge}=\{\left(x,y\right)~|~ 0\leq x \leq \mu,
~0\leq y \leq \mu g\left(x/\mu\right)+2\ge\}
~,\]
\[  {\bf 1}_{ G^-_{\mu,\ge}}
={\bf 1}_{
 0\leq x \leq \mu,~0\leq y \leq \mu g\left(x/\mu\right)-2\ge }
-{\bf 1}_{0\leq x  \leq \mu,~  \mu g\left(x/\mu\right)-2\ge \leq y
  \leq 0}  
~.\]
\begin{figure}[hbtp]
  \begin{center}
    \leavevmode
\input{domain-G.pstex_t}
       \caption{the domains $ G^\pm_{\mu,\ge}$}
    \label{fig:G}
  \end{center}
\end{figure}

For each $m\notin \mu G$ with $m\in {\cal R}$,
  $B\left(m,\ge\right)\cap G^-_{\mu,\ge}  =\emptyset$, 
hence $(\1_{G^-_{\mu,\ge}}\star\gr_\ge)(m)=0$ while  $\forall (x,y)\in \R^2,~
0\leq( \1_{G^-_{\mu,\ge}}\star\gr_\ge)(x,y)\leq 1$.
Similarly,
 for each $m\in \mu G \cap {\cal R}$, $B\left(m,\ge\right)\subset
G^+_{\mu,\ge}$ and  $(\1_{G^+_{\mu,\ge}}\star\gr_\ge)(m)=1$.
Hence,
\[ N_\ge^-  \left(\mu \right)\leq N\left(\mu\right) \leq N_\ge^+
 \left(\mu \right)~.\] 
We will  apply Poisson summation
formula and  use  estimates on the  Fourier transform
of $\1_{G^\pm_{\mu,\ge}}$.
Let us denote by 
\[ \Phi^\pm_{\mu, \ge} \left(\xi,\eta\right)=\int_{\R^2} \chi_0 (x,y)
{\bf 1}_{G^\pm_{\mu,\ge}}(x,y)e^{i(x\xi+y\eta)}dxdy  \]
 the Fourier
 transforms of  $\chi_0 $-times the  characteristic function
of $G^\pm_{\mu, \ge}$.

The Poisson summation formula applied to the sum
(\ref{equ:neps}) gives
\begin{equation}\label{equ:PSF}
  N_\ge^\pm  \left(\mu \right)=\int_{\bf \R^2}
\chi_0(x,y){\bf 1}_{G^\pm_{\mu,\ge}}(x,y) dxdy 
+ \sum _{\left( p,q \right) \in  \Z^2 \setminus 0}
\hat{\rho}\left( 2\pi \ge \left(p,q \right) \right)
\Phi^\pm_{\mu,\ge}\left(2\pi p,2\pi q\right)
e^{-2\pi i\beta q}~.\end{equation} 

We need to evaluate $ \Phi^\pm_{\mu,\ge}$. 
We use Green-Riemann formula in order to get  integrals on the
boundaries.
We have the following formulas:
\begin{lemm}If $\chi _0$ is a smooth classical symbol of degree $0$ and
 \[ \ga =\frac{i}{\eta} \left( \chi _0 (x,y) +\frac{i}{\eta}\pa _y
\chi _0 (x,y) - \frac{1}{\eta^2}\pa _{yy}\chi _0 (x,y) \right)
e^{i(x\xi +y\eta )} dx ~,\]
then
 \[ d\ga =\chi_0 (x,y)e^{i(x\xi +y\eta )} dx \wedge dy  
+ O(\eta ^{-3})\chi_1(x,y) dx \wedge dy ~,\]
where $\chi_1(x,y)\in L^1( dx dy)$.

A similar results holds for 
 \[ \gb =\frac{1}{i\xi} \left( \chi _0 (x,y) +\frac{i}{\xi}\pa _x
\chi _0 (x,y) - \frac{1}{\xi^2}\pa _{xx}\chi _0 (x,y) \right)
e^{i(x\xi +y\eta )} dy~.\]
\end{lemm}

If $|\eta | \leq C |\xi| $, we use
\[ e^{i\left(x\xi +y \eta\right)}\chi_0(x,y) dx\wedge dy =
 d\gb + 0(1/\xi^3)\chi_2 dx\wedge dy  ~,\]
while, if  $|\xi | \leq C |\eta |$, we use
 \[ e^{i\left(x\xi +y \eta\right)}\chi _0 (x,y)dx\wedge dy =
 d\ga + 0(1/\eta^3)\chi_1 dx\wedge dy  ~.\]
We have to estimate the integrals
\[ \int _{\pa G^\pm _{\mu,\ge}} e^{i\xi\left(x +\nu y\right)}
  \chi_0\left(x,y\right) dy ~,\] where $\nu =\eta /\xi$ is bounded
(and similar integrals with $\chi_0$ replaced by the derivatives of 
$\chi _0$ which are  symbols of $<0$
degrees)
and
\[ \int _{\pa G^\pm _{\mu,\ge}} e^{i\eta\left(y +\nu x\right)}
  \chi_0\left(x,y\right) dx ~,\]
 where $\nu =\xi/\eta $ is  bounded.
We use the upper bounds given in Appendix A for the different 
parts of the boundaries, using the parametrization of the graph of $y=g(x)$
by $x(t)=1-t^2 f(t),~y(t)=t^3$  for $0\leq t \leq t_0$. 
For example, the main part of the integral on the curved part 
of $\pa G^+_{\mu,\ge }$ of $\ga $ is
\[\frac{i}{\eta} \int_0^\infty \chi_0 (x(t),y(t))e^{i\mu \eta (y(t)+
\nu x(t))}x'(t) dt ~,\]
to which we apply estimate given in Lemma \ref{lemm:integral}.
This gives:
\begin{lemm}  \label{lemm:estim} The following estimates hold:
 \begin{itemize}
\item \[\Phi^\pm_{\mu, \ge} \left(0,0\right)=
\mu ^2\int _G \chi\left(y/x\right) dx dy + O\left(\mu ^{2/3}\right)\]
\item
For $1 \leq |p| \leq C |q|$,
\[\Phi^\pm_{\mu, \ge} \left(2\pi p, 2 \pi q\right)=
 O \left( \frac{\mu^2}{\left(1+ \mu \| \left(p,q\right) \|\right)^{3/2}}
+ \frac{1}{|p q| }\right) \] 
\item
For $p=0, ~q\ne 0$,
\[\Phi^\pm_{\mu, \ge} \left(0, 2 \pi q\right)= \frac{\mu i }{2\pi q}
   + O \left( \frac{\mu^2}{\left(1+ \mu |q|\right)^{3/2}}\right)
 \]
 \item
For $|q|\leq C |p|$,
\[ \Phi^\pm_{\mu, \ge} \left(2\pi p, 2 \pi q\right)=
 O\left( \frac{\mu^2}{\left(1+ \mu \| \left(p,q\right) \|\right)^{3/2}}
+ \frac{1}{\mu ^{1/3}|p | }\right) ~.\]
\end{itemize}
\end{lemm}
Let us prove for example the estimate 
of $ \Phi^\pm_{\mu, \ge} \left(0, 2 \pi q\right)$.
The corresponding integral on the boundary splits into 2 parts
$\int _0^\mu .. dx -\int _0^{t_0}.. dt $.
The first part gives the first term. The second part gives, up to
constants,
$J=q^{-1}\int _0^{t_0}{\rm exp}(2\pi i q \mu t^3) \chi_0 (\mu x(t),\mu y(t))
\mu x'(t) dt $  which is bounded by
$0(q^{-1}\mu (q\mu)^{-1/2})$ using the estimates 
of Lemma \ref{lemm:integral}.

We need also the classical formula:
\begin{lemm}
For $0<\beta <1$, we have
\[i \sum_{q\in \Z \setminus 0} e^{-2\pi i \beta q}/q =
2\pi \left( \beta -\ha \right)~. \]
\end{lemm}

Theorem \ref{theo:lpp} follows then from the previous Lemmas
and simple evaluations of the sums in the Poisson summation
formula (\ref{equ:PSF});  using 
the fact that the Fourier transform of $\gr $ is rapidly decaying,
 we need  the bounds:
\begin{lemm}
We have:
\[\mu^2 \sum _{(p,q)\in \Z^2\setminus 0}(1+ \mu \| (p,q) \|)^{-3/2}
 (1+ \mu^{-1/3} \| (p,q) \|)^{-N} =O\left( \mu ^{2/3} \right)~,\]
\[ \sum _{1\leq |p|\leq C |q|}|pq|^{-1}(1+ \mu^{-1/3} \| (p,q)
\|)^{-N}
=0 \left((\log \mu)^2\right) ~,\]
\[ \mu^2 \sum _{q\ne 0}|(1+\mu |q|)^{-3/2}(1+ \mu^{-1/3}|q|)^{-N}
=0 \left( \mu^\ha\right) ~,\]

\[\mu^{-1/3} \sum _{1\leq |q|\leq C |p|}|p|^{-1}(1+ \mu^{-1/3} \| (p,q)
\|)^{-N}
=0 (1) ~.\]
\end{lemm}
Let us check the first upper bound, the others are similar.
The first  sum is bounded by
\[ C \mu^\ha \sum  _{(p,q)\in \Z^2\setminus 0}\| (p,q) \|^{-3/2}
(1+ \mu^{-1/3} \| (p,q) \|)^{-3/2}\] 
which is of the same order as the integral
\[ \mu^\ha  \int_0^\infty \frac{rdr}{r^{3/2}(1+\mu^{-1/3}r)^N}~.\]

\hfill$\square $

\section{Spectrum of the disk as a lattice point problem}
\label{sec:lpp/disk}
Our goal is to prove the following result:
\begin{theo} \label{theo:compare}
\[ N_{\rm disk}(\mu )=N_{D}(\mu)+ O\left( \mu^{2/3} \right)~.\] 
\end{theo}
This will complete  the proof of Theorem \ref{theo:main}.

\begin{demo}
The estimate splits into  3 parts: the inner part and the 2 boundary parts.
We choose a function $\chi \in C_o^\infty (]-1, 1[,[0,1])$ which is
$\equiv 1 $ in some large interval $[-1+c,1-c]$ and split the two  numbers
to compare as
\[ N_{\rm disk}(\mu )= N_{\rm disk}^1 (\mu )+
 N_{\rm disk}^2 (\mu )+ N_{\rm disk}^3 (\mu )=\]
\[= \sum_{x_k(n)\leq \mu} \chi (k/n)
+  \sum_{n\geq 0,~x_k(n)\leq \mu} (1-\chi (k/n))
+ \sum_{n< 0,~x_k(n)\leq \mu} (1-\chi (k/n)) ~,\]
and 
\[ N_{D}(\mu)= N_{D}^1(\mu)+ N_{D}^2(\mu)+ N_{D}^3(\mu)=\]
\[ =\sum_{(n,k+\max(0,-n)-1/4)\in \mu D } \chi (k/n)
+  \sum_{n\geq 0,~(n,k-1/4)\in \mu D } (1-\chi (k/n))
+ \sum_{n< 0,~(n,k+|n|-1/4)\in \mu D } (1-\chi (k/n)) ~.\]
We will compare the first terms (the inner parts)
in both decompositions and the second
terms (the boundary parts).
The third ones are similar to the second ones.

{\it The inner part:}
The zeros $x_k(n)$ 
 are given uniformly in any
 domain $x_k(n)> (1+c) |n|$
with $c>0$,   by
\[ x_k(n)= \left\{ 
\begin{array}{l} F(n,k-\frac{1}{4} )+O(1/(1+k+n))~{\rm if}~ n\geq 0 \\
 F(n,k+|n|-\frac{1}{4} )+O(1/(1+k+|n|)) ~{\rm if}~ n<     0 
\end{array}
\right.  \]
This is a consequence of the stationary phase expansion applied
to the integral representations (\ref{equ:bessel}) of Bessel functions
(see \cite{CGJ}).
Using the fact that when $F(n,k+\max(0,-n)-1/4)$ is close to $\mu$,
$|n|+k$ is of the same order as $\mu$, we get
\[ N_D^1\left(\mu -\frac{C}{\mu}\right)\leq N^1_{\rm disk}(\mu)
\leq  N_D^1\left(\mu +\frac{C}{\mu}\right)~.\]
It  follows then from the Van der Corput's 
 remainder estimate $O\left(\mu ^{2/3}\right)$
for the smooth strictly convex  lattice point problems that
\[  N_{\rm disk}^1 (\mu )- N_{D}^1(\mu)=O\left(\mu^{2/3}\right)~.\]

{\it The boundary parts:}
due to the fact that the zeros of $J_n$ and $J_{-n}$ are the
same, we discuss only the case $n>0$. We are in a domain where 
 $x_k(n)< (1+C)n$.
Let us denote by $t_k$ the $k$-th zero of the Airy function, we have
(see \cite{AS})
\[ t_k= \left[ \frac{3\pi}{2}\left( k-\frac{1}{4}\right) \right]^{2/3} 
+\ge(k) ~,\] 
with $\ge(k)= O\left(k^{-1}\right)$. 
From \cite{CGJ}, we have the following equation for the zeros
$x$
of the Bessel function $J_n(x) $ :
\[  {\rm Ai}\left(|x|^{2/3}\rho \left(u\right)\right)
+ b\left(u,x\right)|x|^{-4/3} {\rm Ai}'\left(|x|^{2/3}\rho
 \left(u\right)\right)= 0 ~,\]
where  $u=\left(n/x\right)-1 <0 $,
 $\rho $ is a smooth germ  of odd diffeomorphism
of $\left(\R,0\right)$ (with $\rho' >0$)
 and $b\left(u,x\right)$ is a smooth  symbol of degree  $0$ in $x$.
 We deduce, using the implicit function theorem and
the asymptotics of the Airy function,  that
\[  x_k\left(n\right)=n\left(1+\psi\left( \frac{t_k}{n^{2/3}}\right)\right)+
\eta(n,k)\]  with 
$\eta(n,k)=O(n^{-1})$,  
$\psi $ smooth, $\psi (0)=0$, $\psi'(0)>0$.
This asymptotics is due to Olver \cite{Ol}. If $(1+C)n> k> (1+c)n >0 $
with $0<c<C $, this
asymptotics matches
with the inner asymptotics via the asymptotics of the $t_k$'s for
 large $k$'s.

Let
 \[ N_k\left(\mu \right):=\# \{ \left(n,k-1/4\right)\in
 \mu D~|~1\leq k \leq Cn \}\]
and 
\[ N'_k\left(\mu \right):=\# \{ x_k(n)\leq \mu   ~|~1\leq k \leq Cn
\}~.\]
We have the
\begin{lemm}\label{lemm:nk}
\[  |N_k\left(\mu \right)-  N'_k\left(\mu \right)|\leq
  N_k\left(\mu +\frac{C}{\mu}\right)-   N_k\left(\mu -\frac{C}{\mu}\right)+
 C\mu^{1/3}k^{-4/3}~.\]
\end{lemm}
By summing the estimate of the previous Lemma w.r. to $k$
 and using the 2-terms asymptotics of
$N_D^2(\mu)$,
we get 
\[N_{\rm disk}^2(\mu )- N_D^2(\mu)
=O\left(\mu ^{2/3}\right)~. \]
\end{demo}

{\it Proof of Lemma \ref{lemm:nk}:}
Let us write  $F(x,y):=x(1+ \psi_1 ( y^{2/3}/x^{2/3}))$; we have
$N_k(\mu)=\# \{ F(n,k-1/4)\leq \mu~ |~~1\leq k \leq Cn \}$
and
$ N'_k\left(\mu \right)=\# \{
  F(n,k-1/4+\ge(k))\leq \mu +\eta (k,n) ~|~1\leq  k \leq Cn \}$,
with $\ge(k)= O(1/k)$ and $\eta (k,n)= O(1/n)$.
We have,
in the range $0< y\leq cx $ with $c$ small enough, 
 $0<a < \pa _x F <b $ and $\pa _y  F= O(x^{1/3}y^{-1/3})$.
The Lemma follows by estimating the cardinal of the sets
\[A_k:=\{ n~|~  \mu  \leq F(n,k-1/4) \leq  \mu +\eta (k,n)+
 C\ge(k)n^{1/3}k^{-1/3}\}
\]
and 
\[A'_k:=\{ n~|~ \mu -\eta (k,n)- C\ge(k)n^{1/3}k^{-1/3} \leq F(n,k-1/4) \leq
\mu \}~. 
\]
We use the fact that if $F(n,k-1/4)$ is close to $\mu $ then
$n\sim \mu $.
We have $A_k\subset B_k \cup C_k$
with 
$B_k:=\{ n~|~ \mu  \leq F(n,k-1/4) \leq
\mu +O(1/\mu) \}$ and
$C_k:=\{ n~|~ \mu  +\eta (k,n) \leq F(n,k-1/4) \leq
\mu +\eta (k,n)+ C\ge(k)n^{1/3}k^{-1/3} \}$.
Using the estimate on $\pa _x F$,
we have $\# C_k =O\left(n^{1/3}k^{-4/3}\right)$

\section{Conclusion and problems}

It would be nice to get similar estimates for other
integrable billiards like a circular annulus.
 The case of ellipse is more difficult  and is due
to Emile  Mathieu \cite{Mathieu}: 
the problem is with the unstable periodic geodesic (the larger
diameter).
We know now a good approximation of the associated eigenvalues
thanks to my works with Bernard  Parisse and \san ~(\cite{C-Pa,C-Vu}).

\section*{Appendix A: Estimation of some integrals}
\label{app:integrals}

We need to get estimates of various integrals corresponding
to part of the boundary of the domains $D^\pm _{\mu,\ge}$
to be defined in Section \ref{sec:lpp}.

Let us first recall the following stationary phase estimate:
\begin{lemm}\label{lemm:stat}
Let $f\in C^\infty ([a,b],\C)$ and $\phi \in C^\infty  ([a,b],\R)$ 
so that $\phi $ has only non degenerate critical points, then, if 
\[ I(\tau):=\int _a^b e^{i\tau \phi (t)}f(t) dt ~,\]
we have  $I(\tau )=O\left( \tau ^{-\ha}\right) $.
 If $\phi $ depends smoothly on some
parameter
$\mu $ so that the non  degeneracy assumption holds for $\mu =\mu_0$, 
the same conclusion is true
uniformly in some interval $|\mu -\mu_0|\leq c$ with $c$ small enough.
\end{lemm}

\subsection*{The curved part}\label{subsec:curved}
These integrals come when evaluating integrals on the curved part
of the domains $D_{\mu,\ge}^\pm $.
\begin{lemm}\label{lemm:integrala}
Let us consider the integral
\[ I_{c,A}(\tau )=\int _0^\infty e^{i\tau \left(\nu t^3 - t^2
    f(t\right) )}
t^2g(t) dt \]
with $g\in C_o^\infty (\R) $, $f\in C^\infty (\R,\R)$
with $f(0)\ne 0$  and $|\nu |\leq \nu _0 <\infty $ with $\nu _0 $
small enough, 
then, as $\tau \ra \infty $,  $I_{c,A}(\tau) =O(\tau^{-\ha})$ if $|\nu
_0|$
is small enough.
\end{lemm}
This is easy using the  stationary phase Lemma \ref{lemm:stat}.
\begin{lemm}\label{lemm:integral}
Let us consider the integral
\[ I_{c,B}(\tau )=\int _0^\infty e^{i\tau \left(t^3 -\nu t^2 f(t\right) )}tg(t) dt \]
with $g\in C_o^\infty (\R) $, $f\in C^\infty(\R,\R)$,
 and $|\nu |\leq \nu _0 <\infty $ with
$\nu_0 $ small enough,
then, as $\tau \ra \infty $,  $I_{c,B}(\tau) =O(\tau^{-\ha})$.
\end{lemm}
This is more difficult because the critical point $t=0$ is degenerate
for $\nu=0$.
We need a 
\begin{defi}
A smooth function $f(t,\alpha)$ is in $S^k$ if all $t$-derivatives are bounded
near $t=\infty$ by $O(t^k)$ uniformly in $\alpha$.
\end{defi}
\begin{demo}
We will first prove the Lemma for $f\equiv 1$.
Let us put $\mu = \nu \tau^{1/3}$.
We consider 2 cases:
\begin{itemize}
\item $|\mu |\leq 1$: let us make the change $t=w\tau^{-1/3}$,
we get
\[ I_{c,B}(\tau)=\tau^{-2/3}\int _0^\infty e^{i\left(w^3 -\mu
  w^2\right)}wg\left(w\tau^{-1/3}\right)dw ~,\]
The critical points of the phase are $0$ and $2\mu /3$.
We split the integral into 2 parts with a smooth partition of unity
$1=h+ (1-h)$ with $h\in C_o^\infty(  [0,w_0[) $ and $h\equiv 1$ on $[0,1]$.
 The part containing the critical
points is $O\left( 1\right)$ using an uniform bound
for the integrand. For the  other part, we introduce
$L=1/\left(3w^2-2\mu w\right)d/dw$ and 
integrate by part several times using the formal transpose  $^tL$
of $L$  and the fact
that $wg\left(w\tau^{-1/3}\right)\in S^1$, so that 
$\left(^t L\right)^N\left(wg\left(w\tau^{-1/3}\right)\right) \in S^{1-2N} $.
If $N\geq 2$, this gives a function which is in $L^1(]w_0,+\infty [)$
uniformly in $\mu $. 
\item $|\mu |\geq 1$:
we put $t=\nu \sigma $ and get
\[ I_{c,B}(\tau)= \nu^2 \int_0^\infty e^{i\mu ^3 \left(\sigma^3 -\sigma^2\right)}
\sigma g\left(\nu \sigma\right)d\sigma ~.\] 
We split the integral smoothly and get for the part containing the
critical
points $O\left(\nu^2/\mu^{3/2}\right)=O\left(\tau^{-1/2}\right)$.
For the other part we use
$K=\left(1/3w^2 -2w\right)d/dw $, $^t K:S^k \ra S^{k-2}$ and
$\sigma g\left(\nu \sigma\right)\in S^1$.
We pick a factor $\mu^{-3}$ for
each integration by parts and 
get $O\left( \nu^2 / \mu ^{3N}\right)= O\left( \tau^{-2/3}\right)$.
\end{itemize}
It is clear enough that the proof still works if $f$ is not constant. 
\end{demo}

\subsection*{The linear parts}\label{subsec:linear}
These integrals come when evaluating integrals on the linear  parts
of the domains $D_{\mu,\ge}^\pm $.

\begin{lemm}\label{linear}
We have, for $|\eta |=O(|\xi|)$,  $I_v\left(\xi,\eta\right)=
\frac{1}{\xi } \int_0^{2\ge}  
e^{iy\eta } dy =O\left(\ge /| \xi |\right)$.

For $\xi=O\left(|\eta|\right)$,  if 
\[ I_h\left(\xi,\eta\right)=\frac{i}{\eta}\int_0^\mu e^{ix\xi} dx ~,\]
then,  for $\xi \ne 0$, 
$I_h\left(\xi,\eta\right)=O\left(1/|\xi \eta|\right)$ for $\xi \ne 0$
and 
$I_h\left(0,\eta\right)=\frac{i\mu }{\eta}$. 
\end{lemm}

\bibliographystyle{plain}

\end{document}